\documentclass[aps,groupedaddress]{revtex4}
\usepackage{amsfonts} % para incluir fonts y
\usepackage{amssymb}  % simbolos de la AMS
\usepackage{graphicx} % standard LaTeX graphics tool
\usepackage{amsmath}  %  paquete matemÃ¡tico
\usepackage{aas_macros}
\usepackage[colorinlistoftodos]{todonotes}

\begin{document}

%\preprint{}

\title{Kicks of magnetized strange quark stars induced by anisotropic emission of neutrinos}


\author{Alejandro Ayala}
\affiliation{Instituto de Ciencias Nucleares, Universidad Nacional Aut\'onoma de M\'exico, Apartado Postal 70-543, CdMx 04510, Mexico}
\affiliation{}
\affiliation{Centre for Theoretical and Mathematical Physics, and Department of Physics, University of Cape Town, Rondebosch 7700, South Africa.}

\author{D. Manreza Paret}
\affiliation{Instituto de Ciencias Nucleares, Universidad Nacional Aut\'onoma de M\'exico, Apartado Postal 70-543, CdMx 04510, Mexico}
\affiliation{}
\affiliation{Facultad de F\'isica FF-Universidad de la Habana, San Lazaro y L. Vedado, 10400 La Habana, Cuba.}

\author{A. P\'erez Mart\'inez}
\affiliation{Instituto de Cibern\'etica Matem\'atica y F\'isica (ICIMAF), Calle E esq a 15 Vedado, 10400 La Habana, Cuba.}

\author{Gabriella Piccinelli}
\affiliation{Centro  Tecnol\'ogico, Facultad de Estudios Superiores Arag\'on, Universidad  Nacional  Aut\'onoma  de  M\'exico, Avenida  Rancho  Seco  S/N,  Col. Impulsora Popular Av\'icola, Nezahualc\'oyotl,  Estado  de  M\'exico  57130,  Mexico.}

\author{Angel S\'anchez}
\affiliation{Departamento de F\'isica, Facultad de Ciencias, Universidad Nacional Aut\'onoma de M\'exico, Apartado Postal 70-542, Ciudad de M\'exico 04510, Mexico.}

\author{Jorge S. Ru\'iz Monta\~no}
\affiliation{Facultad de Ciencias de la Tierra y el Espacio, Universidad Aut\'onoma de Sinaloa.
Blvd. de la Americas y Av. Universitarios S/N, Ciudad Universitaria, Culiac\'an, Sinaloa 80007, Mexico.}

\date{\today}

\begin{abstract}
We study the anisotropic neutrino emission from the core of neutron stars induced by the star's magnetic field. We model the core as made out of a magnetized ideal gas of strange quark matter and implement the conditions for stellar equilibrium in this environment. The calculation is performed without resorting to analytical simplifications and for temperature, density and magnetic field values corresponding to typical conditions for a neutron star's evolution. The anisotropic neutrino emission produces a rocket effect that contributes to the star's kick velocity. We find that the computed values for the kick velocity lie within the range of the observed values,  reaching velocities of the order of $\sim1000$ km s$^{-1}$ for magnetic fields between $10^{15}-10^{18}$ G  and radii of 20 to 5 km, respectively.
\end{abstract}

%\keywords{}

\maketitle

\section{Introduction}\label{intro}

One of the great challenges faced by modern astrophysics is the understanding of the internal composition of compact objects such as Neutron Stars (NS). The interest is also stimulated by current efforts to map the phase diagram of nuclear matter in the high baryon density, low temperature domain, where heavy-ion experiments have little or no chance to contribute. It is speculated that at baryon densities several times higher than normal nuclear density, strongly interacting matter undergoes a phase transition into quark matter. A nice argument based on the large $N_c$ expansion~\cite{McLerran:2007qj} shows that when the quark chemical potential exceeds the constituent quark mass, the increase in pressure produces a phase where chiral symmetry is restored, though quarks are still confined and baryons are parity doubled. A further increase in density produces that quarks roam on top of a baryon Fermi surface to form the so called {\it quarkyonic matter}. For sufficiently high densities, this matter can contain even strange quarks to form the so called {\it strange quark matter} that could arrange itself into a quark color superconductor phase such as the color-flavor locked (CFL) phase.

Compact objects can also sustain high intensity magnetic fields, whose surface strength ranges form $10^8$ G for millisecond pulsars and $10^{12}$ G for radio pulsars~\cite{lyne2006pulsar}, up to $10^{15}$ G for magnetars~\cite{Ibrahim:2002zw}. It has been estimated that these surface field intensities imply that in the NS core the magnetic field can reach a strength as high as $10^{18}-10^{19}$~G \cite{1991ApJ...383..745L,Isayev:2012sv}.

Recently, the sophisticated techniques to detect gravitational waves developed by LIGO and other observatories~\cite{2016PhRvL.116f1102A, 2017PhRvL.119p1101A}, have opened a new window to astronomy. As an example one can mention the recent detection of gravitational waves corresponding to the merging of two neutron stars~\cite{2017PhRvL.119p1101A, GBM:2017lvd}. These ground observatories, together with satellite ones, such as the {\it Neutron star Interior Composition ExploreR} (NICER) Mission \footnote{https://heasarc.gsfc.nasa.gov/docs/nicer/, https://heasarc.gsfc.nasa.gov/docs/nicer/} will soon provide constraining measurements of NS properties with uncertainties below 10\%~\cite{2016ApJ...832...92O}. This privileged modern scenario is encouraging to also carry out deeper theoretical studies on the composition of neutron stars.

The presence of magnetic fields in the interior of compact objects affects not only the equation of state (EoS) but also the solutions to Einstein's equations. For the former, the magnetic field influences the micro-physics, by means of the Landau quantization of the energy levels experienced by moving charged particles. This quantization produces the anisotropy of the energy-momentum tensor and the splitting of pressure into parallel and perpendicular (to the magnetic field) components. Effects on the structure equations come from this anisotropy in the energy-momentum tensor.
In previous works, the anisotropic effect of magnetic fields on the EoS for white dwarfs (WD), neutron stars and quark stars (QS)~\cite{Chaichian:1999gd,Aurora2003EPJC,PhysRevC.77.015807} as well as for the evolution of the early universe~\cite{2016GReGr..48....7D} have been studied. The possibility that unpaired magnetized quark matter be in a superconductor state, and in particular in the CFL phase, was explored in Refs.~\cite{Felipe2009JPhG,2011EPJA...47....1G}. For WD and QS, calculations have been carried out to analyze the anisotropic structure of equilibrium equations~\cite{1674-4527-15-10-1735,1674-4527-15-7-975}.

Another phenomenon that may be influenced by the presence of a magnetic field in the interior of compact objects is the translational velocity observed for some pulsars. This velocity corresponds to a peculiar compact object's motion with respect to the surrounding stars and to their progenitors. Data corresponding to the motion of 233 pulsars has been collected in Ref.~\cite{2005MNRAS.360..974H} where velocities as high as $1000\ $km\,s$^{-1}$ as well as mean velocities for young pulsar of $400\ $km\,s$^{-1}$ are reported. These NS translational velocities are commonly referred to as {\it pulsar kick velocities}.

Several mechanisms have been  proposed  to explain these kicks, the most relevant ones can be found in Ref.~\cite{2001ApJ...549.1111L}. Depending on the time of their appearance, whether at birth or during the NS evolution, the kicks have been classified into natal or post-natal, respectively. Since the kick corresponds to a NS's asymmetric motion, all the proposed models rely on some kind of asymmetric velocity producing mechanism. Among these we can mention the asymmetric matter ejection due to hydrodynamical perturbations during the core collapse and the supernova explosion (natal kick) \cite{1970ApJ...160L..91G}; the electromagnetic rocket effect produced by electromagnetic radiation along the NS spin axis from an off-centered rotating magnetic dipole (post-natal) \cite{1975ApJ...201..447H} and the asymmetric emission of neutrinos from the core of a NS in the presence of strong magnetic fields, coming from changing-flavor charged currents \cite{1985SvAL...11..123D}. The latter mechanism seems to be the most natural one since the emission of neutrinos is the most efficient for the NS cooling. This could be responsible for either a natal or post-natal kick, depending on the main reaction that is taking place in the core.

In this work, we study the asymmetric emission of neutrinos from a NS core transiting into magnetized strange quark (MSQ) matter.  This process could lead to the formation of either a hybrid star (namely, a core of MSQ matter with a crust made out of nuclear matter), or a bare MSQ star. For the purpose of our study, the final stage of the NS does not matter, thus we refer indistinctly to one or the other case as a MSQ.

The mechanism of asymmetric emission of neutrinos has already been explored in Refs.~\cite{2008JPhG...35a4062S,Sagert:2007as} where it was shown that, when ignoring neutrino quark scattering and for typical values of temperature, density and magnetic field strength in a NS core, it is possible to achieve kick velocities of order 1000 km s$^{-1}$. These large velocities receive corrections due to the smaller neutrino mean free path when neutrino interactions are included. Such interactions produce that the neutrino motion within the core quickly reaches a diffusive stage which translates into a reduced anisotropic neutrino motion.  When considering these corrections, the largest kick velocities that can be obtained are of order 100 km s$^{-1}$ even when color superconductivity effects are included. Corrections induced from non-Fermi liquid behavior on the neutrino mean free path and emissivity beyond the leading order have also been considered in Ref.~\cite{Adhya:2012sq, Adhya:2013ima}. Nevertheless, other important ingredients for neutrino propagation within the core have not yet been explored. These include the possible magnetic field induced decrease of the neutrino coupling with the surrounding plasma. A similar effect for QCD matter has been found, albeit for high temperature and not for high density, for the coupling constant in QCD and effective QCD models~\cite{Ayala:2017ucc,Ayala:2015bgv,Ayala:2014uua,Ayala:2014gwa}. This reduction of the coupling strength is linked to the so called {\it inverse magnetic catalysis} phenomenon, observed by lattice QCD calculations~\cite{Bali:2012zg,Bali:2011qj,Bali:2014kia}. In addition, the effects on the neutrino scattering arising from the anisotropic pressures are yet to be explored. Since when considered, these ingredients may play a significant role for the properties of neutrino propagation in dense magnetized media, here we first revisit the underlying main mechanism responsible for neutrino emission during the evolution of a neutron star. Following Ref.~\cite{Sagert:2007as}, we study a more realistic scenario by imposing the stellar equilibrium ($\beta$ equilibrium, charge neutrality and baryon number conservation) conditions in the core of the NS treating it as a magnetized gas of strange quark matter (SQM) and by taking into account the magnetic field dependence in the chemical potential and temperature of all of the thermodynamical quantities involved. The calculation is performed without resorting to analytical simplifications and for temperature, density and magnetic field values corresponding to typical conditions for a neutron star's evolution.

The paper is organized as follows: In Sec.~\ref{sec2} we start by recalling the expression that relates the relevant thermodynamical quantities to the kick velocity. In Sec.~\ref{sec3} we give a brief sketch of the thermodynamical properties of magnetized fermions. In Sec.~\ref{sec4} we compute the electron polarization in the presence of a magnetic field and finite chemical potentials. In Sec.~\ref{sec5} we present our results for the kick velocities when the stellar equilibrium equations are considered. Finally, in Sec.~\ref{sec6} we discuss the plausibility of our approach in light of effects that can influence the neutrino mean free path within the NS core and we give our conclusions.

\section{Pulsar kick velocity} \label{sec2}

We consider a scenario where the neutrino emission mainly comes from the NS core, where the density is taken high enough so that the significant degrees of freedom are the $u$, $d$ and $s$ quarks and neutrino (antineutrino) emission comes from beta decay. Although the emitted particles can travel in all directions, a rocket effect appears when an imbalance in the emission of neutrinos with a momentum component in one versus the opposite direction exists. This direction is provided by the magnetic field, which is taken to point in a single direction, at least for distances of the size of the neutron star’s core.  Since beta decay is a parity violating process, the relevant quantity entering the calculation of this kick velocity  is the electron spin polarization asymmetry, defined as the ratio between the difference and the sum of the number of spins polarized electrons in the directions along and opposite to the magnetic field.  Conservation of momentum and angular momentum in a parity violating process require that the emitted neutrino-positron (antineutrino-electron) spin polarization are correlated with their direction of motion projected along or opposite to the direction provided by the magnetic field. Therefore the number of neutrinos (antineutrinos) emitted in one or the opposite direction can simply be computed from their spin asymmetry.
Notice that  the phase space factors coming from the whole beta decay process cancel in the ratio that leads to the calculation of the spin electron polarization asymmetry, leaving only the expressions for the number of particles (electrons) with one and the other spin which  can be computed accounting for a thermal environment and in the presence of a magnetic field.  The efficiency for the process is a complicated function of temperature, chemical potential and magnetic field strength. It is 100\% efficient only for extremely large values of the magnetic field. However, as we show, for finite field strengths, the polarization grows as the intensity of the magnetic fields increases. Notice that for not too extreme values one can achieve a non-vanishing asymmetry and that, in fact, a rough estimation indicates that an asymmetry of order 3 \% could be sufficient to power a kick of order 1000 km s$^{-1}$ for a 1.4 solar mass neutron star (see Ref.~\cite{2008JPhG...35a4062S}). We also assume that the angle between the magnetic field direction and the axis of rotation is small for the mechanism to be more efficient, as is discussed in Ref.~\cite{2007ApJ...660.1357N}.

The produced kick velocity of the NS can be computed as \cite{2008JPhG...35a4062S, Sagert:2007as}
\begin{equation}\label{vel_kick}
dv=\frac{\chi }{M_{NS}}\frac{4}{3}\pi R^3\epsilon dt,
\end{equation}
where  $M_{NS}$ and $R$ are the NS mass and radius, $\epsilon$ is the neutrino emissivity and $\chi$ is the electron spin polarization, which  determines the fraction of neutrinos asymmetrically emitted as we mentioned before.

When the emissivity changes with temperature, the cooling equation can be used, namely,
\begin{equation}\label{vel_kick3}
-\epsilon=\frac{dU}{dt}=\frac{dU}{dT}\frac{dT}{dt}=C_v\frac{dT}{dt},
\end{equation}
where $U$ is the internal energy density and  $C_v$ is the heat capacity. Therefore, the kick velocity is given by

\begin{equation}\label{velchi1}
v=-\frac{1 }{M_{NS}}\frac{4}{3}\pi R^3\int_{T_i}^{T_f}\chi\, C_vdT.
\end{equation}
This velocity can be written in the following form
\begin{equation}\label{velchi2}
v=-803.925\, \frac{\text{km}}{\text{s}}\left( \frac{1.4M_\odot}{M_{NS}}\right)\left( \frac{R}{10 \,\text{km}}\right)^3 \left( \frac{I}{\text{MeV}\,\text{fm}^{-3}}\right),
\end{equation}
where
\begin{equation}\label{vel2}
I=\int_{T_i}^{T_f}\chi\, C_vdT.
\end{equation}

The integral  $I$ is a function  of $C_v$ and $\chi$ and both depend on magnetic field, chemical potential and temperature. These quantities are also derived from thermodynamical properties of the system which we proceed to calculate.

\section{Thermodynamics of magnetized fermions at low temperatures}\label{sec3}

In order to find $C_v$ and $\chi$ we need to compute the thermodynamical potential in the presence of a magnetic field, including the contribution from all species subject to the magnetic field effects.

Let us consider the core of the NS made out of MSQM,  namely a gas composed of quarks $u$, $d$ and $s$ and electrons $e$ in the presence of a magnetic field.

The thermodynamical potential for a fermion species $f$ in a constant magnetic field directed along the $\hat{z}$ direction reads
\begin{equation}\label{Thermo-Potential-2}
\Omega_f(B,\mu_f,T)=-\frac{e_fd_fB}{2\pi^2}\int_{0}^\infty dp_3\sum_{l=0}^{\infty}(2-\delta_{l0})%\nonumber\\ &\times&
\left[E_{lf}+\frac{1}{\beta}\ln \left(1+e^{-\beta(E_{lf}-\mu_f)}\right)\left(1+e^{-\beta(E_{lf}+\mu_f)}\right)\right],
\end{equation}
where
\begin{equation}\label{Disp-Rel}
E_{lf}=\sqrt{p_3^2+2|e_f|Bl+m_f^2}, \quad l=\nu+\frac{1}{2}+\frac{s}{2} ,\quad\nu=0,1,2,\ldots,\quad s=\pm 1,
\end{equation}
$e_f$ and $m_f$ are the fermion charge and mass respectively, $l$ is the Landau level, $s$ is the particle's spin and $p_3$ is the momentum along the magnetic field $\textbf{B}$.

From Eq.~(\ref{Thermo-Potential-2}), we can find the expressions for all relevant thermodynamical quantities such as the particle number density, the entropy and the heat capacity, given by
\begin{equation}
n_f=-\frac{\partial\Omega_f}{\partial\mu_f} \quad, \quad S_f=-\frac{\partial \Omega_f}{\partial T} \quad\mbox{and}\quad  C_{vf}=T\frac{\partial S_f}{\partial T},
\end{equation}
respectively.

In particular, the explicit expression for particle density $n_f$ and heat capacity $C_{vf}$ read as
\begin{equation}\label{ecn}
n_f=\frac{d_fm_f^3}{2\pi^2}b_f\sum_{l=0}^{\infty}(2-\delta_{l0})\int_0^\infty dx_3 \frac{1}{e^{(\frac{m_f}{T}\sqrt{x_3^2+1+2lb_f}-x_f)}+1},
\end{equation}
\begin{equation}\label{Cv}
C_{vf}=\frac{d_fm^2}{4\pi^2 T^2}b_f\int_{0}^\infty dp_3\sum_{l=0}^{\infty}(2-\delta_{l0})\frac{(E_{lf}-\mu_f)^2}{[1+\cosh{\frac{E_{lf}-\mu_f}{T}}]},
\end{equation}
where  $x_3=p_3/m_f$ is the dimensionless momentum along the magnetic field and $b_f=B/B_f^c$, being $B_f^c=m_f^2/e_f$ the critical magnetic field. As a reference we take the electron critical field value: $B_c^e=m_e^2/e_e=4.41\times 10^{13}$ G and will use as short notation  $b_e\equiv b=B/B_e^c$.

We can also obtain the well known expression for the $T=0$ limit of the thermodynamical potential given by \cite{PhysRevC.77.015807}
\begin{equation}\label{Thermo-Potential-52}
\Omega_f(B,\mu,0) = -\frac{d_fm_f^4}{4\pi^2}b_f\sum_{l=0}^{l_{max}}(2-\delta_{l0})\left[x_f p_{lf}-\varepsilon_{lf}^2\ln\left(\frac{x_f+p_{lf}}{\varepsilon_{lf}}\right)\right],
\end{equation}
where  $x_f=\mu_f/m_f$ is the dimensionless chemical potential,  $p_{lf}=(x_f^2-\varepsilon_{lf}^2)^{1/2}$, $\varepsilon_{lf}=[1+2lb_f]^{1/2}$ and the maximum number of Landau levels is  $l_{max}=I[(x_f^2-1)/(2b_f)]$, with $I[\zeta]$ denoting the integer part of $\zeta$ number.

Since NS  chemical potential values associated to the density are higher than their temperatures ($\mu \gg T$), the approximation in Eq.~(\ref{Thermo-Potential-52}) is good enough to construct an EoS. However, when we are interested in studying temperature dependent phenomena, it is important to properly account for the temperature dependence. This is the case of the present study since we want to  describe a post-natal  kick appearing during the cooling stage of the NS. In this stage, the temperature varies in the range of $10$~MeV to $0.1$~MeV. Thus, the low temperature limit can only be considered as a leading order approximation to the problem. To account for next to leading order corrections, we follow the general procedure described in appendix B of Ref.~\cite{2001NuPhB.612..492D}. In this manner, we obtain the low temperature expansion for the thermodynamical potential, given by
\begin{equation}
\Omega (B,\mu,T) = \Omega_{0}(B,\mu) + \frac{\pi^{2}}{6} T^{2} \Omega^{''}_{0} (B,\mu)  +  \frac{7 \pi^{4}}{360} T^{4} \Omega^{(4)}_{0} (B,\mu) + \ldots\ ,
\label{exprrefapB}
\end{equation}
where the low temperature condition $\mu_f-m_f\varepsilon_{lf}\gg T\geq0$ is used.

Working up to second order corrections in $T$, Eq.~(\ref{exprrefapB}) reads explicitly as
\begin{equation}\label{prins}
\Omega_f(B,\mu,T)=-\frac{d_fm_f^4}{4\pi^2}b_f\left\lbrace \sum_{l=0}^{l_{max}}(2-\delta_{l0})\left[x_f p_{lf}-\varepsilon_{lf}^2\ln\left(\frac{x_f+p_{lf}}{\varepsilon_{lf}}\right)\right]%\right. \\ \left.
+\sum_{l=0}^{l_{max}-1}(2-\delta_{l0})\left(  \frac{\pi^2}{3}\frac{x_f}{p_{lf}}\frac{T^2}{m_f^2}\right) \right\rbrace,
\end{equation}
where the second summation is up to $l_\text{max}-1$ to formally avoid divergences in the analytical expression. The numerical results however, are not affected by this condition since what is missing is just a numerically small term for the whole sum.

From Eq.~(\ref{prins}), the low temperature expressions for the thermodynamical quantities are given explicitly by
\begin{equation}\label{lowT}
n_f(B,\mu,T)=\frac{d_fm_f^3}{2\pi^2}b_f\left[ \sum_{l=0}^{l_\text{max}}(2-\delta_{l0}) p_{lf}-
\frac{\pi^2}{6}\frac{T^2}{m_f^2}\sum_{l=0}^{l_\text{max}-1}(2-\delta_{l0}) \frac{\varepsilon_{lf}^2}{p_{lf}^3}  \right],
\end{equation}
and
\begin{equation}\label{lowTC2v}
C_{vf}=\frac{d_fm_f^2}{6}b_f T\sum_{l=0}^{l_\text{max}-1}(2-\delta_{l0})\frac{x_f}{p_{lf}}.
\end{equation}

In addition to the approximate expression for the heat capacity at low temperature, given by Eq.~(\ref{lowTC2v}), we also calculate the exact expression for the heat capacity from Eq.~(\ref{Cv}). Both expressions will be used in the calculation of the kick velocity, given by the integral $I$ in Eq.~(\ref{vel2}). The comparison between both results is shown in Sec.~\ref{sec5}.

\subsection{Electron polarization in a magnetic field}\label{sec4}

Now we need to compute the second ingredient to determine the kick velocity, namely, the polarization of emitted electrons.
As we mentioned before, the fraction of neutrinos asymmetrically emitted is equal to the number of spin polarized electrons in the presence of the magnetic field.
Due to the constrains imposed by the relation among the quantum numbers, Eq.~(\ref{Disp-Rel}), the electrons in $l \neq 0$ with spin up get aligned parallel ($s=+1$) to the magnetic field and those with spin down anti-parallel ($s=-1$) to it. Meanwhile, the electrons in the Lowest Landau Level (LLL)  can only align antiparallel to the direction of the magnetic field.

Then the fraction of spin polarized electrons could be calculated as
\begin{equation}\label{chi1}
\chi=\frac{n_- - n_+}{n_- +n_+},
\end{equation}
where $n_{\pm}$ are the number densities corresponding to spin oriented parallel and anti--parallel to the magnetic field direction. In this way, $n_{\pm}$ are given by
\begin{equation}\label{ecnp}
n_+=\frac{d_em_e^3}{2\pi^2}b\sum_{\nu=1}^{\infty}\int_0^\infty dx_3 \frac{1}{e^{(\frac{m_e}{T}\sqrt{x_3^2+1+2\nu b}-x_e)}+1},
\end{equation}
\begin{equation}\label{ecnn}
n_-=\frac{d_em_e^3}{2\pi^2}b\sum_{\nu=0}^{\infty}\int_0^\infty dx_3 \frac{1}{e^{(\frac{m_e}{T}\sqrt{x_3^2+1+2\nu b}-x_e)}+1}.
\end{equation}

Here we  used the relation between $l$, $\nu$ and $s$, given by Eq.~(\ref{Disp-Rel}) and  expressed the summation over $l$ in terms of the summation over $\nu$. For this purpose, note that the change in the summation is carried out considering $\sum_{l=0}^{\infty}(2-\delta_{l0})\rightarrow \sum_{s=\pm 1}\sum_{\nu=0}^{\infty}$.

From Eqs.~(\ref{chi1}), (\ref{ecnp}) and (\ref{ecnn}) we can compute the dependence of $\chi$ with the parameters $B,\, T,\, \text{and}\, \mu$, obtaining
\begin{equation}\label{chi2}
\chi=\left\lbrace 1+ \frac{2
	\sum\limits_{\nu=1}^{\infty}\int_0^\infty dx_3 \frac{1}{e^{(\frac{m_e}{T}\sqrt{x_3^2+1+2\nu b}-x_e)}+1}}{\int_0^\infty dx_3 \frac{1}{e^{(\frac{m_e}{T}\sqrt{x_3^2+1}-x_e)}+1}} \right\rbrace ^{-1}.
\end{equation}

This equation is exact and will be computed numerically.  However, an analytical approximation for low temperatures can be written as
\begin{equation}\label{lowTchi}
\chi=\left\lbrace 1+\frac{2\sum\limits_{\nu=1}^{l_\text{max}}p_{le}- \frac{\pi^2}{3}\frac{T^2}{m_e^2}\sum\limits_{\nu=1}^{l_\text{max}-1}\frac{\varepsilon_{lf}^2}{p_{lf}^3} }{p_{0e}-\frac{\pi^2}{6}\frac{T^2}{m^2}\frac{1}{p_{0e}^{3}}}  \right\rbrace ^{-1}.
\end{equation}

\begin{figure}[ht!]
	\centering
	\includegraphics[width=0.45\linewidth]{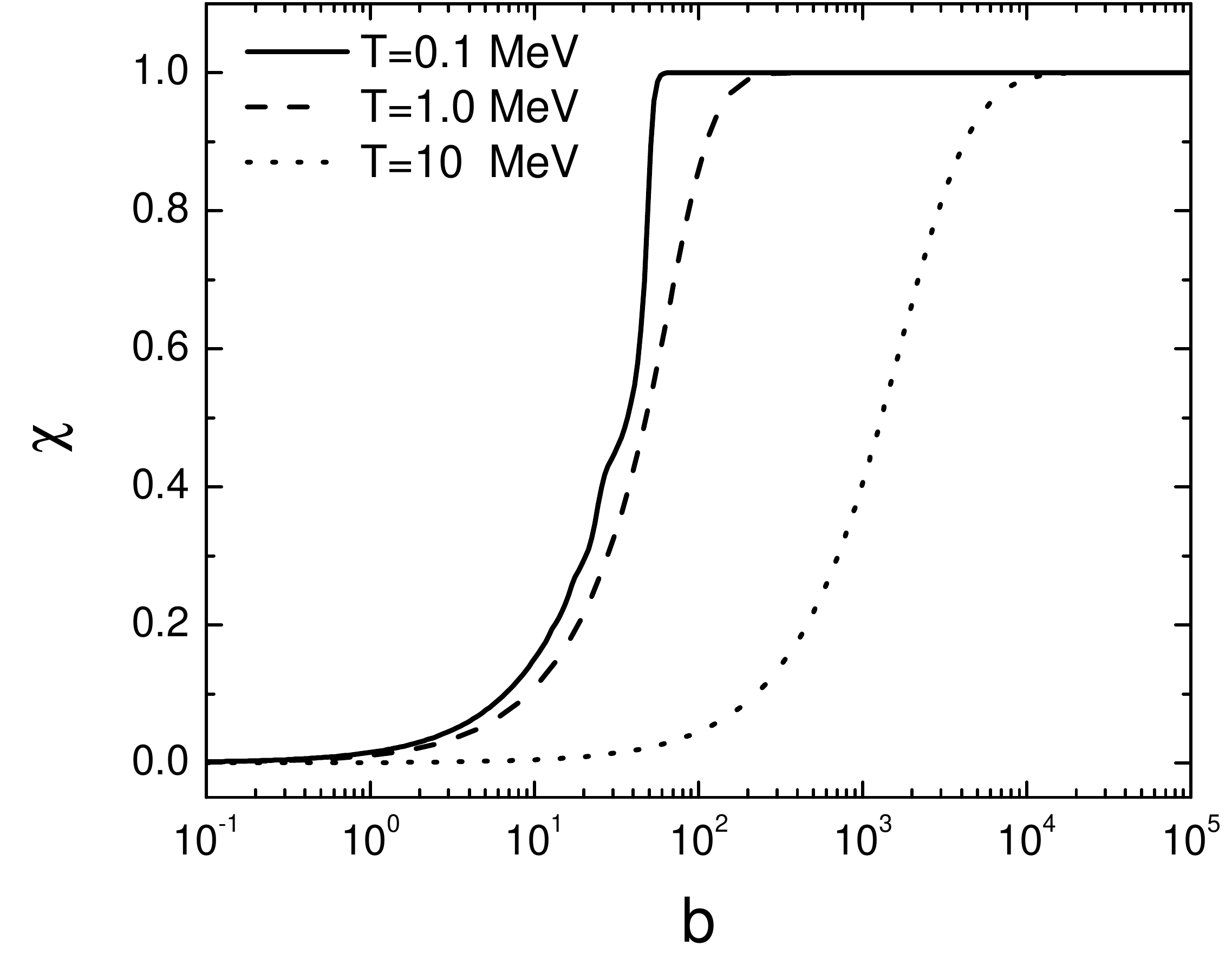}
    \includegraphics[width=0.45\linewidth]{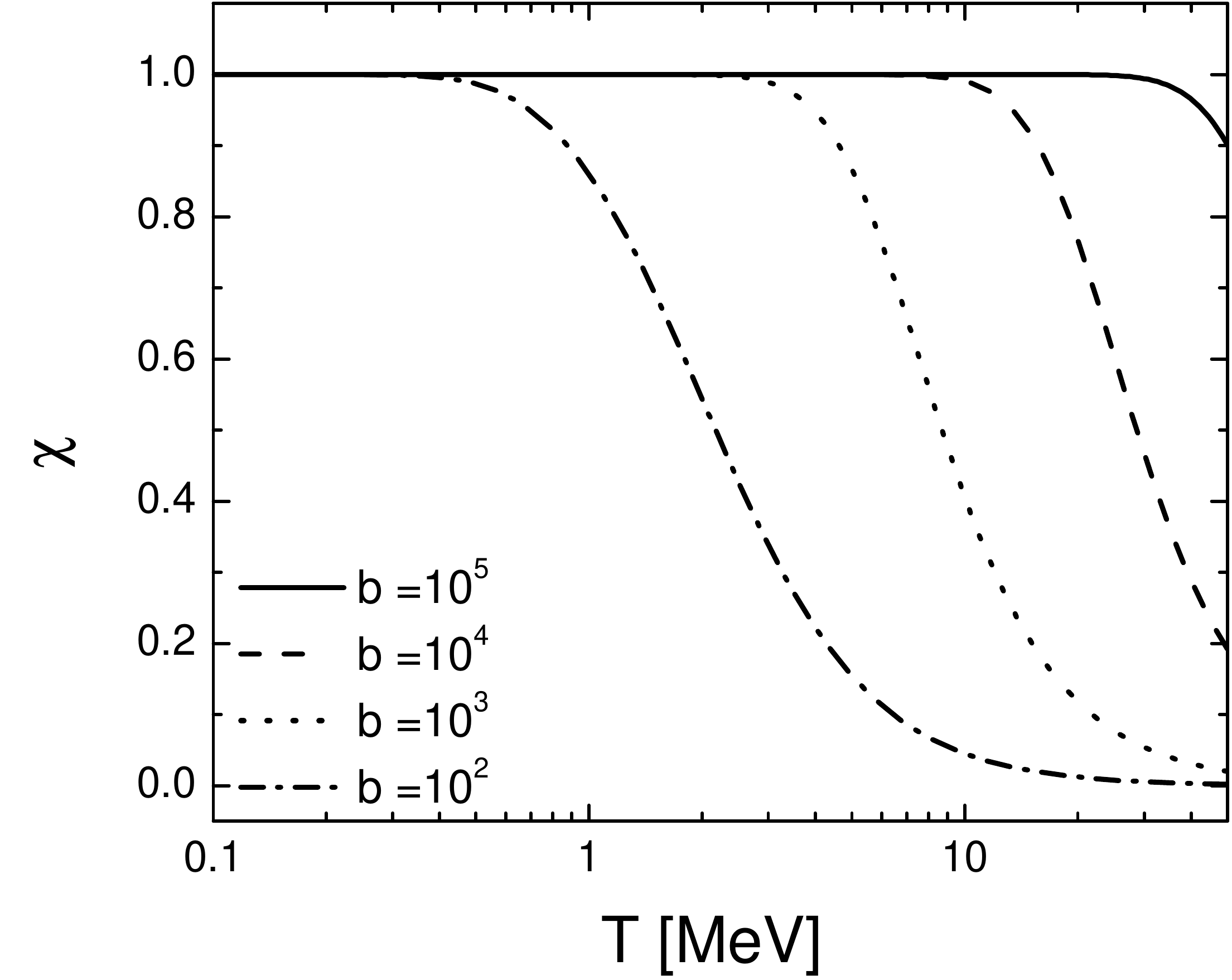}
\caption{ Polarization $\chi$, for a fixed chemical potential of $x_e=10$, as a function of the magnetic field, for different values of the temperature (left panel), and as a function of temperature for several values of the magnetic field (right panel).}
\label{figxiB}
\end{figure}

\begin{figure}[ht!]
	\centering
	\includegraphics[width=0.45\linewidth]{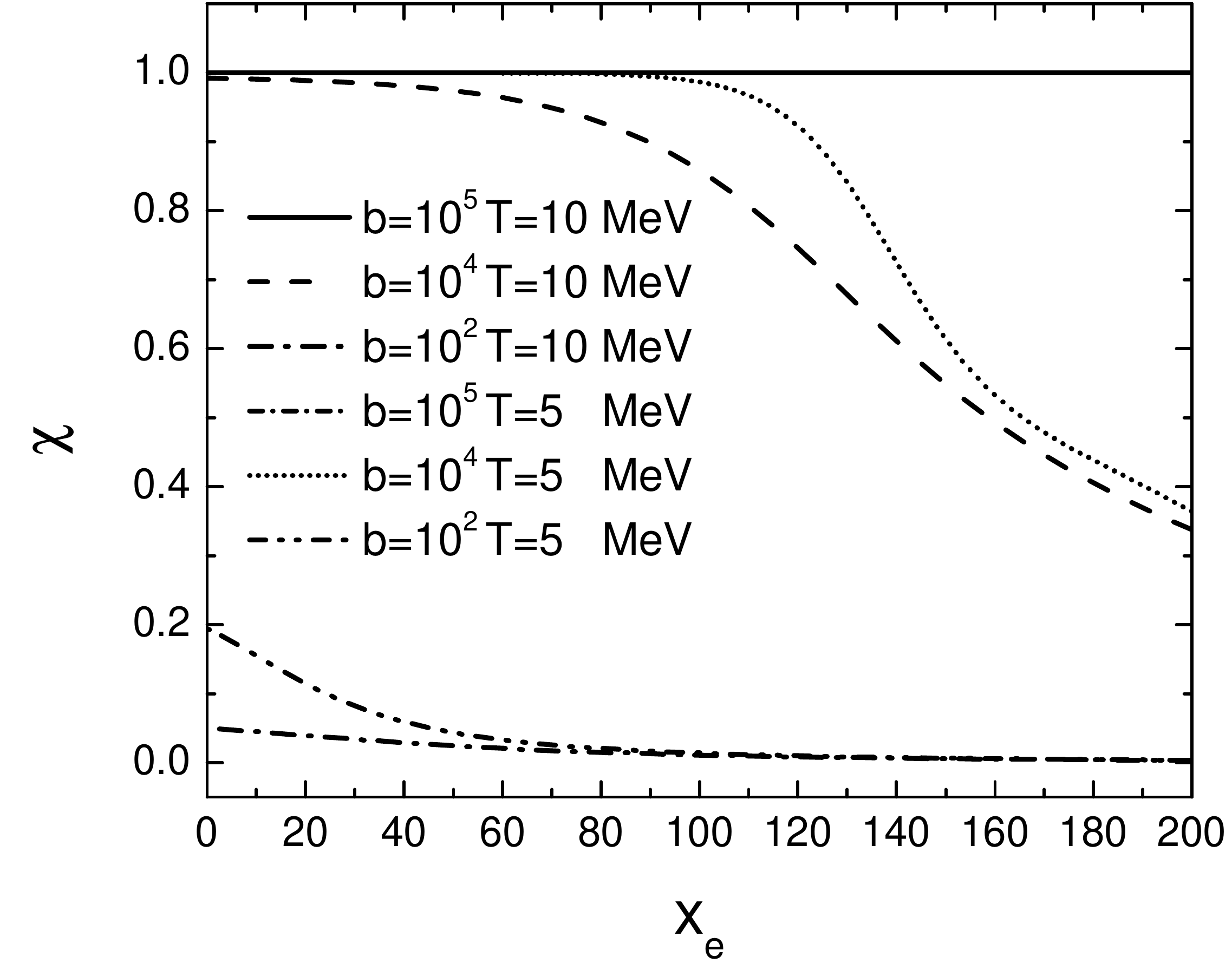}
	\caption{Polarization $\chi$ as function of the chemical potential for several values of the magnetic field and two fixed temperatures.}
\label{figximu}
\end{figure}

 The behavior of the polarization as a function of the chemical potential and temperature is shown in Figs.~\ref{figxiB} and~\ref{figximu}.  The left panel of Fig.~\ref{figxiB} shows that the polarization grows when the magnetic field  increases while the right panel shows that the effect of increasing the temperature is to inhibit the increase of the polarization. Also, notice the typical effect of Landau number transitions for low temperatures can be appreciated. This is similar to the well-known Haas-van Alphen oscillations for the magnetization.

 Fig.~\ref{figximu} shows how the polarization decreases with the chemical potential and the temperature meanwhile  increases with the magnetic field.
\vfill

\begin{figure}[ht!]
	\centering
	\includegraphics[width=0.45\linewidth]{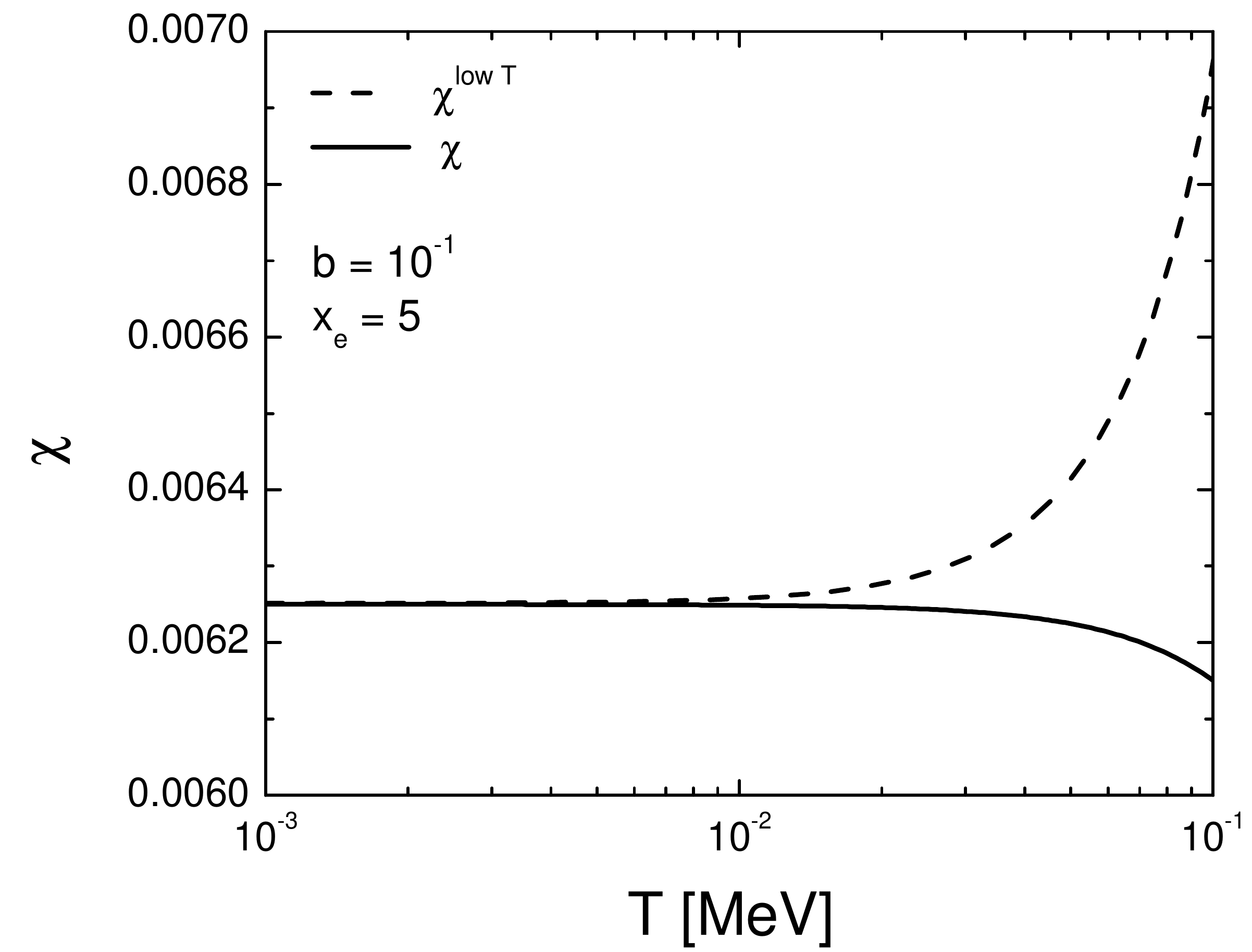}
	\caption{Comparison between the exact and the approximate results for the polarization $\chi$ as function of the temperature for fixed chemical potential of $x_e=5$ and magnetic field of $b=10^{-1}$.  Notice the large discrepancy between these two results that start to be noticeable for temperatures around $10^{-2}$ MeV.}
\label{figxilowT}
\end{figure}

In Fig.~\ref{figxilowT} we compare the numerical and  analytical results (Eq.~(\ref{lowTchi})) in the low-temperature regime. As the temperature decreases, the value of $\chi$ tends to its zero-temperature limit. The tendency of the graphs is different when the temperature increases, since the approximation grows meanwhile the exact value decreases.

\begin{table}[ht!]%
	\caption{Summary of analytical solutions for electron polarization.}{\tabcolsep=2pt%
		\begin{tabular}{|l|c|}
			\hline
			Case & polarization  approximation $(\chi)$ \\
			\hline
			$m \gg T \gg \sqrt{2eB}, \mu_e=0$ & $\chi^W=\frac{1}{2}\frac{m_e}{T}b$ \\ 	
			$\sqrt{2eB} \gg T \gg m_e, \mu_e$&$\chi^S=\left(1-\frac{4}{\text{ln}(2)}\sqrt{\frac{\pi T}{2m_e\sqrt{2b}}}e^{-m_e\sqrt{2b}/T} \right)$\\
			\hline
	\end{tabular}}
\label{table_pol}
\end{table}
\begin{figure}[ht!]
	\centering
	\includegraphics[width=0.45\linewidth]{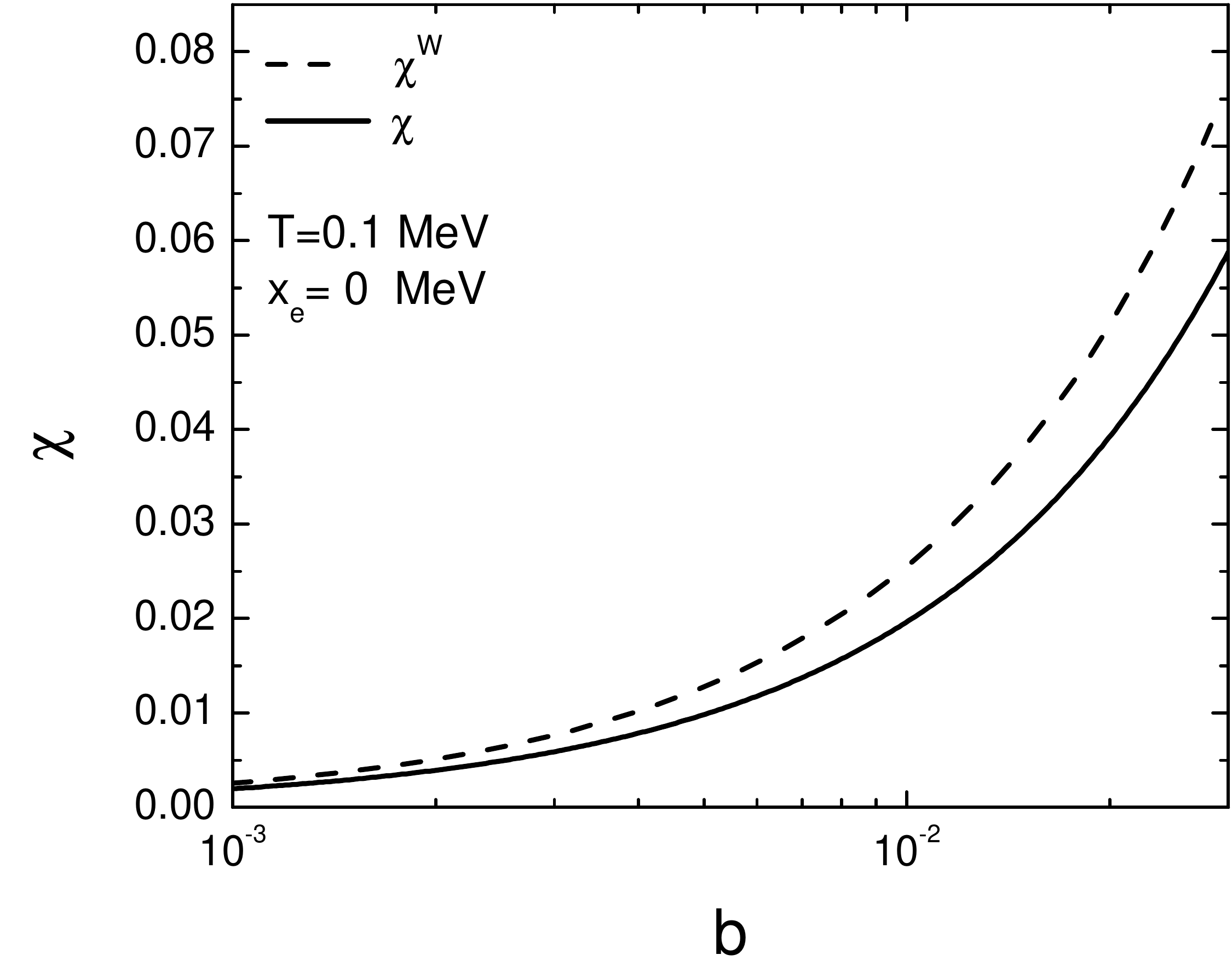}
	\includegraphics[width=0.45\linewidth]{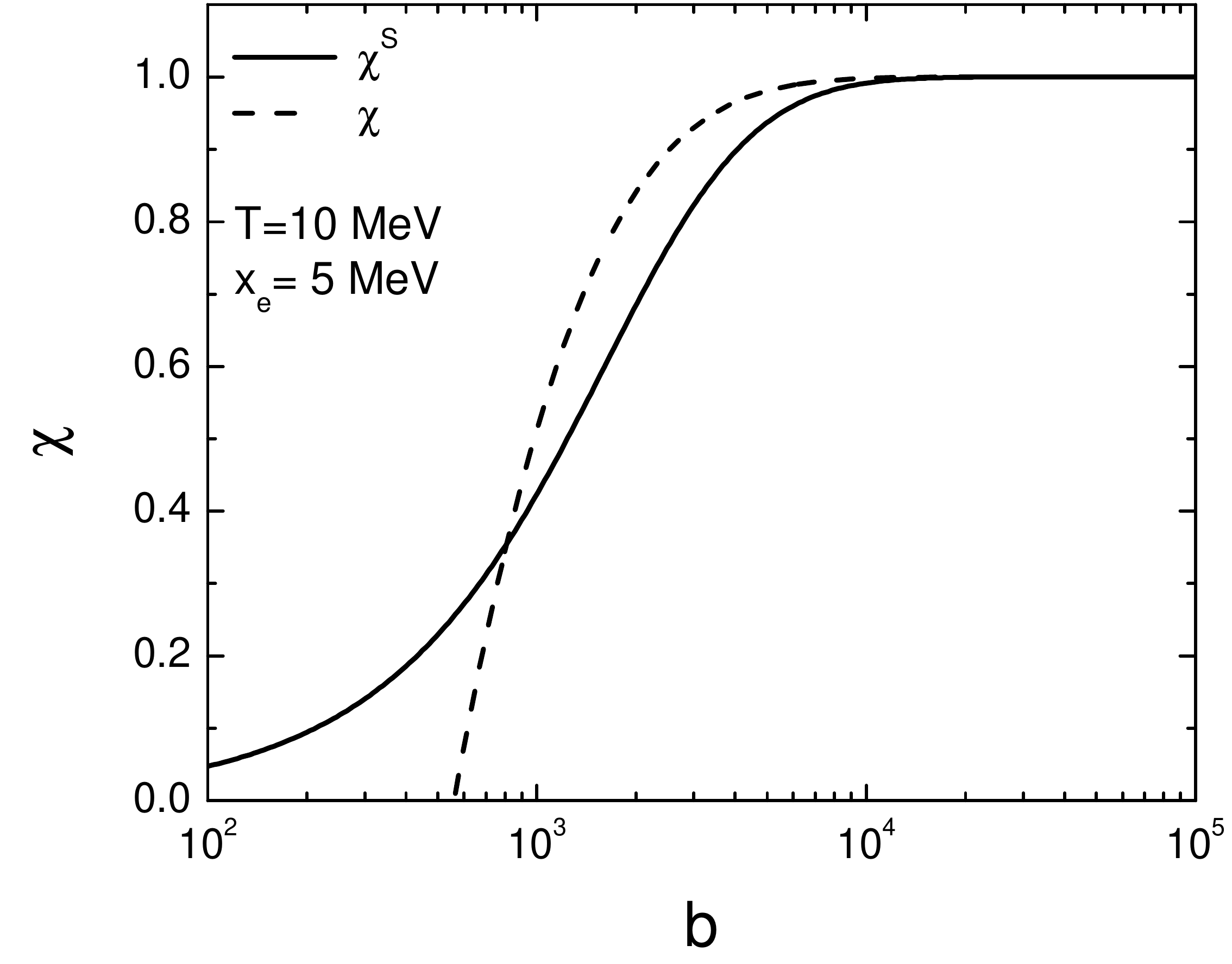}
	\caption{Polarization of electrons $\chi$ as a function of the magnetic field for a fixed value of chemical potential and temperature in the weak field approximation (left panel) and as a  function of the magnetic field in the strong field approximation (right panel).}
\label{figlowstrongB}
\end{figure}
{Figure~\ref{figlowstrongB} shows the behavior of $\chi$ with the magnetic field for exact and for the limit of weak ($\chi^W$) and strong magnetic field ($\chi^S$), whose analytical expressions found in Ref.\cite{Sagert:2007as} are shown in Table I. In left panel we depict the numerical and analytical results for weak magnetic fields. The analytical result corresponds to the first case of Table~\ref{table_pol}. The numerical and analytical results coincide only for weak magnetic fields. The right panel shows the numerical and analytical results corresponding to the second case of Table~\ref{table_pol}. Both results approach the value $\chi = 1$ in the limit of a very strong magnetic field. This is the only regime where both results coincide. Indeed, for fields as high as $b = 10^2-10^3$, both graphs already present very different behaviors, even the approximation becomes negative for magnetic fields around $b=5\times10^2$. This indicates that one has to be careful with the range of values where the strong field approximation can be applied.

\subsection{Stellar equilibrium and numerical results}\label{sec5}
Going back to the kick velocities computation, recall that we require to solve the integral in Eq.~(\ref{vel2}), considering the dependence of $\chi$ and $C_v$ on the magnetic field, the chemical potential and temperature inside the neutron star.  As we are assuming that the kicks are of post-natal origin, we also require to impose the conditions that exist in the core of neutron stars  which are determined by the $\beta$ decay equilibrium equations among the quark species
\begin{subequations}\label{consNb1}
		\begin{eqnarray}
		d&\rightarrow& u+e+\bar{\nu}_e, \,\,\, \,\,\, u+e\rightarrow d+\nu_e,\\
		s&\rightarrow& u+e+\bar{\nu}_e, \,\,\, \,\,\,	u+d\rightarrow u+s.
		\end{eqnarray}
\end{subequations}

Equations~(\ref{consNb}) also represent the charge neutrality and baryon number conservation. All together these equations are referred to as the stellar equilibrium conditions. Their solutions provide the chemical potentials dependence on the temperature. To impose these conditions we have to solve the system of equations
\begin{subequations}\label{consNb}
	\begin{eqnarray}
	\mu_u+\mu_e-\mu_d=0 \,\,\,, \,\,\, \mu_d-\mu_s&=&0,\\
	2n_u-n_d-n_s-3n_e&=&0, \\
	n_u+n_d+n_s-3 n_{B}&=&0,
	\end{eqnarray}
\end{subequations}
where  the chemical potential of neutrinos have been neglected,  $n_f$ are given by Eq. (\ref{ecn}) and $n_{B}$ is the baryon density.

In Fig.~\ref{figmu1} we show the solution (chemical potentials of the fermion species as a function of temperature) of the system of Eqs.~(\ref{consNb}) for a constant central density and a magnetic field of $b=10^5$. We can see a very slight change in the chemical potential with the temperature for all the quark species. The changes in the electron chemical potential are more significant, as it can be appreciated in Fig. \ref{figmu2}.

\begin{figure}[ht!]
	\centering
	\includegraphics[width=0.45\linewidth]{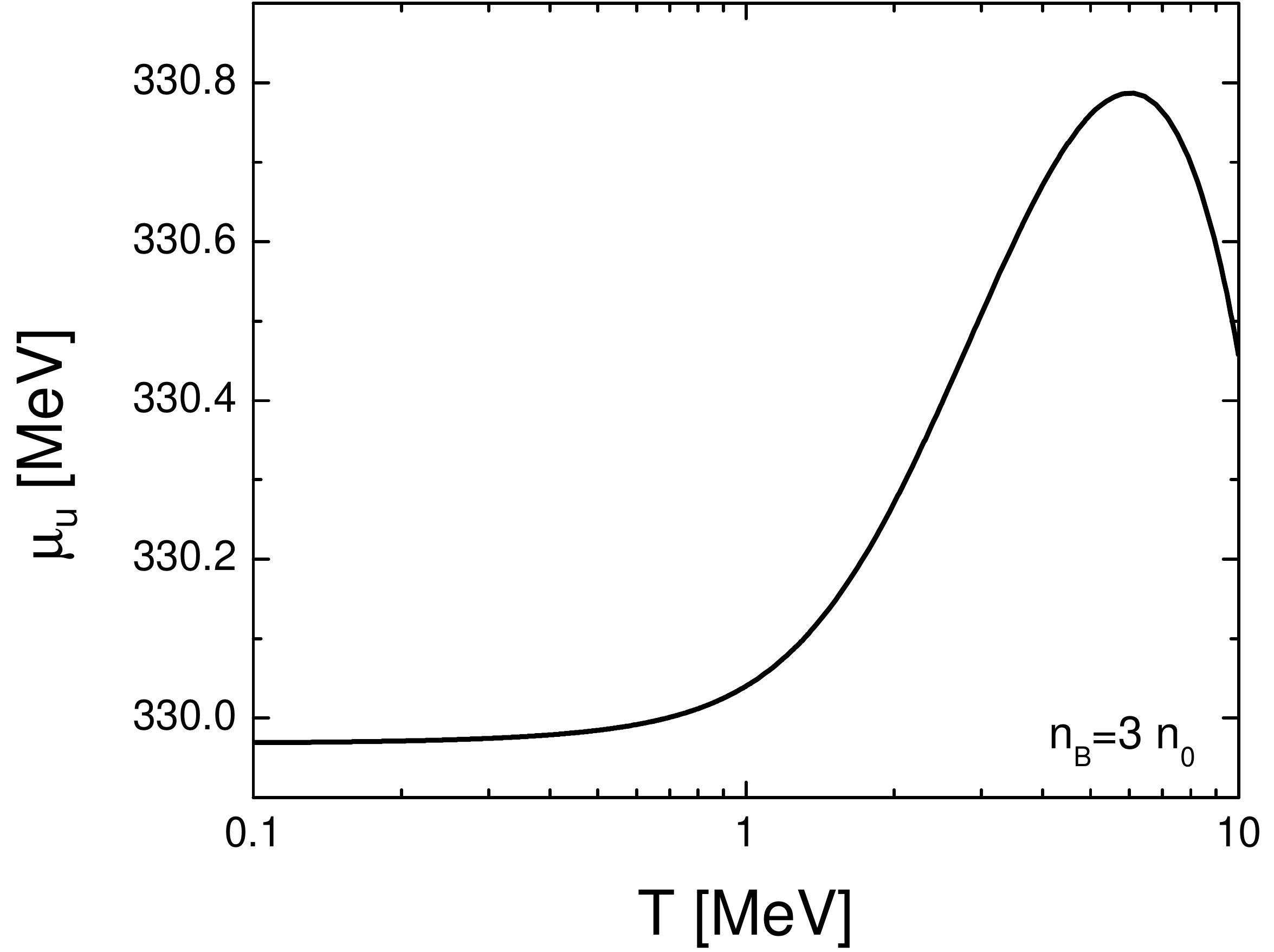}
	\includegraphics[width=0.45\linewidth]{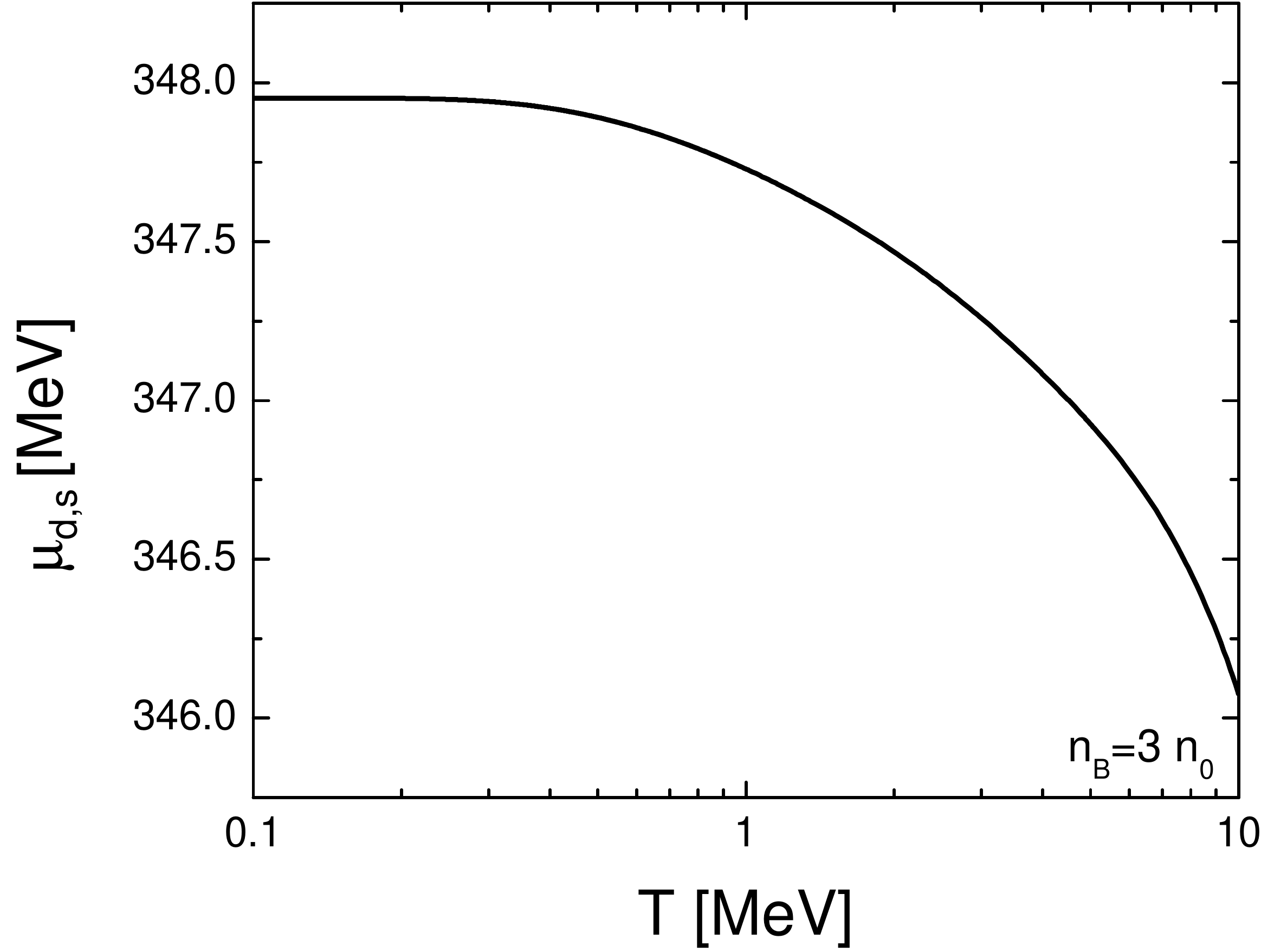}
	\caption{Up (left panel), down and strange (right panel) quark chemical potential as a function of temperature for a fixed value of the central density and  magnetic field $b=10^5$}. \label{figmu1}
\end{figure}
\begin{figure}[ht!]
	\centering
	\includegraphics[width=0.45\linewidth]{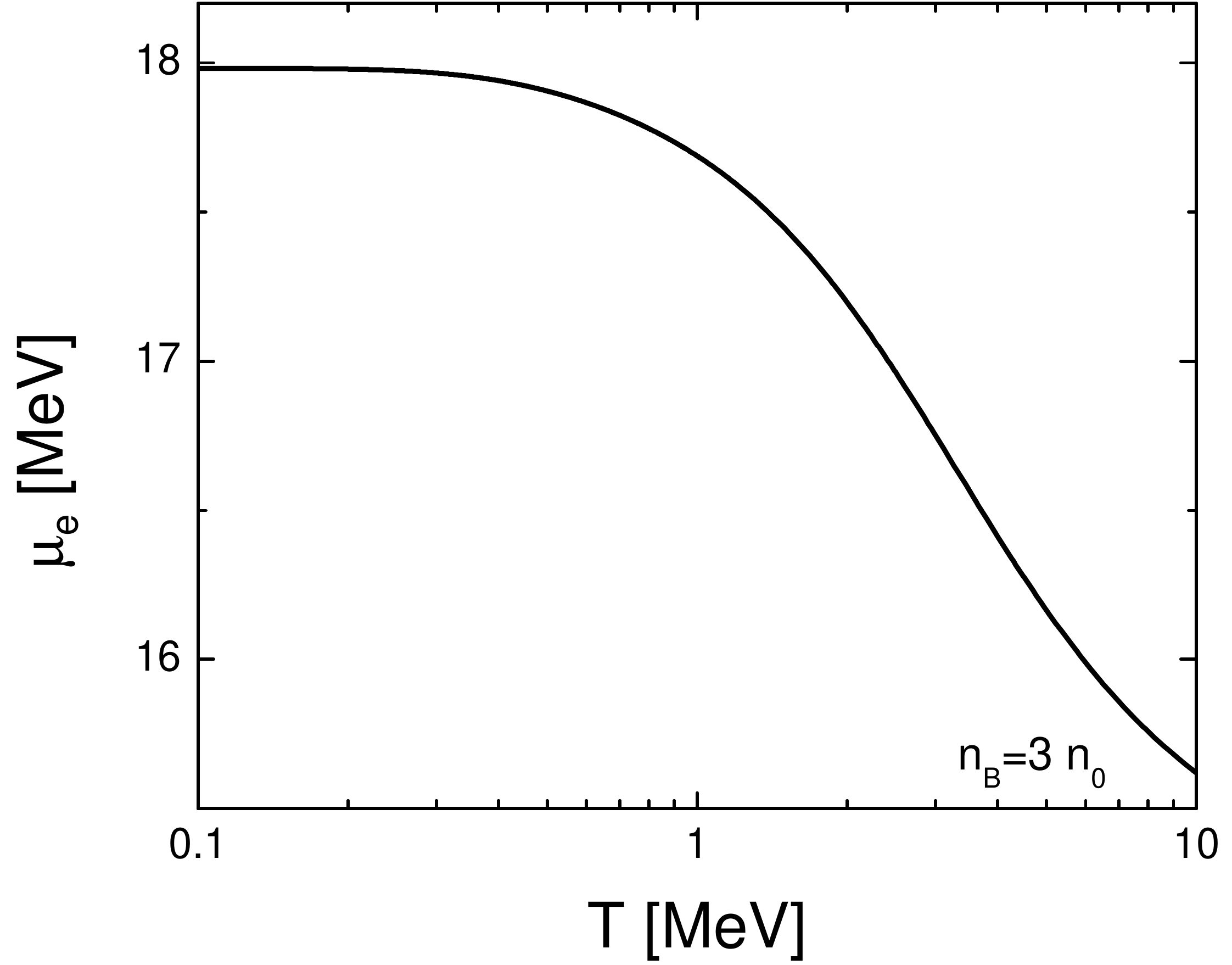}
	\caption{ Electron chemical potential as a function of the temperature for a fixed value of the central density and  magnetic field.}
\label{figmu2}
\end{figure}

In  Fig.~\ref{fig3} we show the specific heat as a function of the temperature for two values of the magnetic field and a fixed baryon density. The figure compares our numerical result for $C_v$  with the ones obtained in Ref.~\cite{Sagert:2007as} where $C_v$ is taken as independent of the magnetic field and the chemical potential and  temperature are fixed to $400$~MeV and $10$~MeV respectively.  Notice that when $C_v$ is computed including its full dependence on the magnetic field, temperature and chemical potential, $C_v$ is consistently smaller than the calculation of Ref.~\cite{Sagert:2007as}. The effect of the number of Landau Levels that contribute can be noticed in the jumps of $C_v$ for certain values of temperature.
\begin{figure}[ht!]
   \centering	
   \includegraphics[width=0.45\linewidth]{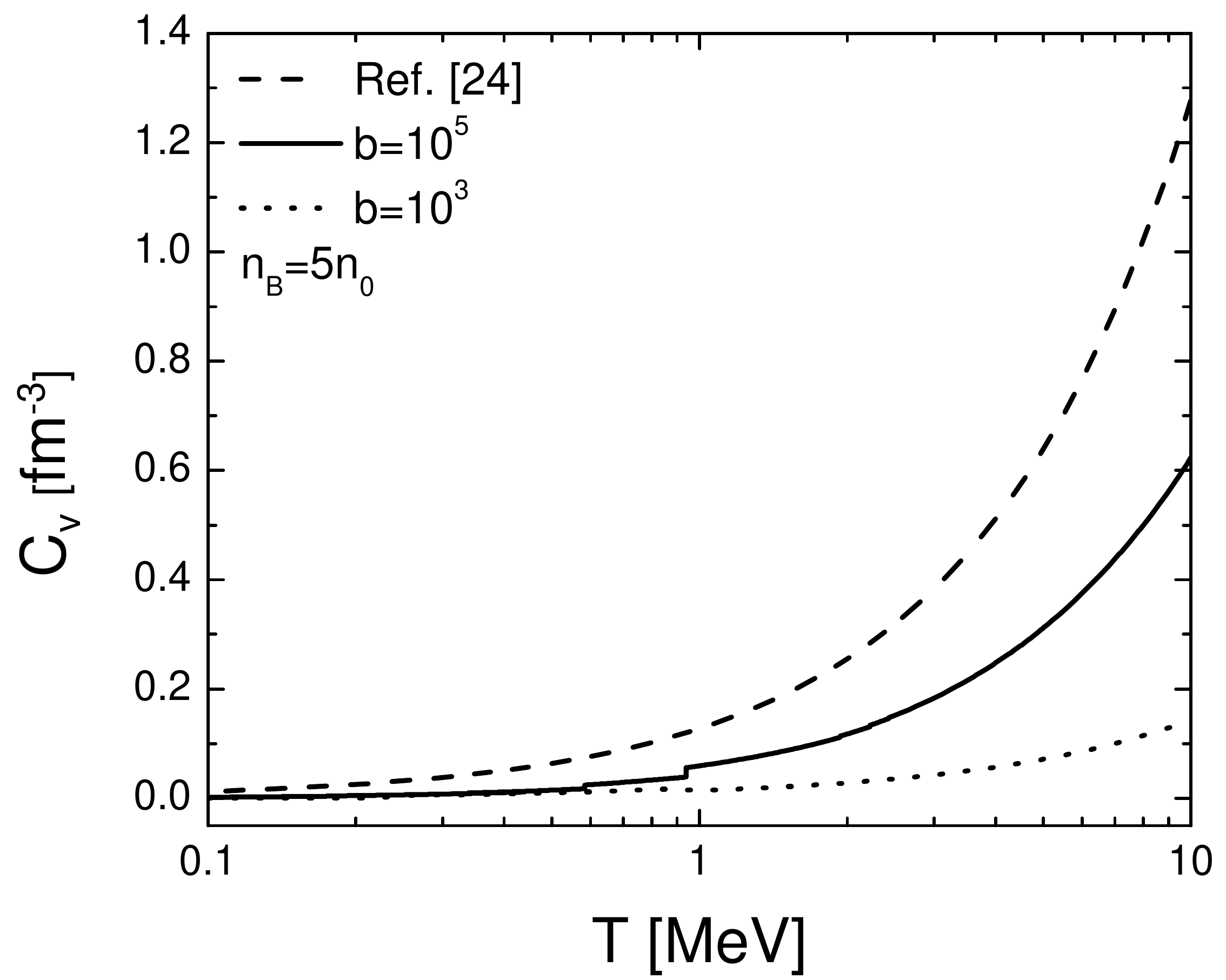}
   \caption {Heat capacity  as a function of the temperature for two values of the magnetic field and a fixed baryon density, compared with the heat capacity obtained in Ref.~\cite{Sagert:2007as}.  \label{fig3}}
\end{figure}

In the left panel of Fig.~\ref{vR1} we show the behavior of the velocity as a function of the neutron star radius for different values of the magnetic field and a fixed central density. In the right panel of Fig.~\ref{vR1} we see how the velocity is affected when the baryon density changes: An increase in the baryon density implies that the stars can reach higher velocities for the same value of magnetic fields and the radius. We compare our numerical calculation with previous results obtained in Ref.~\cite{Sagert:2007as}, where  $\chi=1$ (which is tantamount to magnetic fields higher than $10^{18}$~G), $C_v$ is taken as independent of the magnetic field, the chemical potential and temperature are fixed to $400$~MeV and $10$~MeV respectively, and the stellar equilibrium conditions are neither accounted for. Notice that  our result approaches that of Ref.~\cite{Sagert:2007as} when the magnetic field increases and also that for the highest values of the magnetic field, the neutron star can reach higher velocities for smaller radii, while for low values of the magnetic field the star would require a larger radius to reach velocities of order $v_{\text{kick}}\sim1000$ km s$^{-1}$. These velocities are obtained for magnetic fields between $10^{15}-10^{18}$ G  and radii between 20 to 5 km, respectively.
\begin{figure}[ht!]
	\centering
	\includegraphics[width=0.45\linewidth]{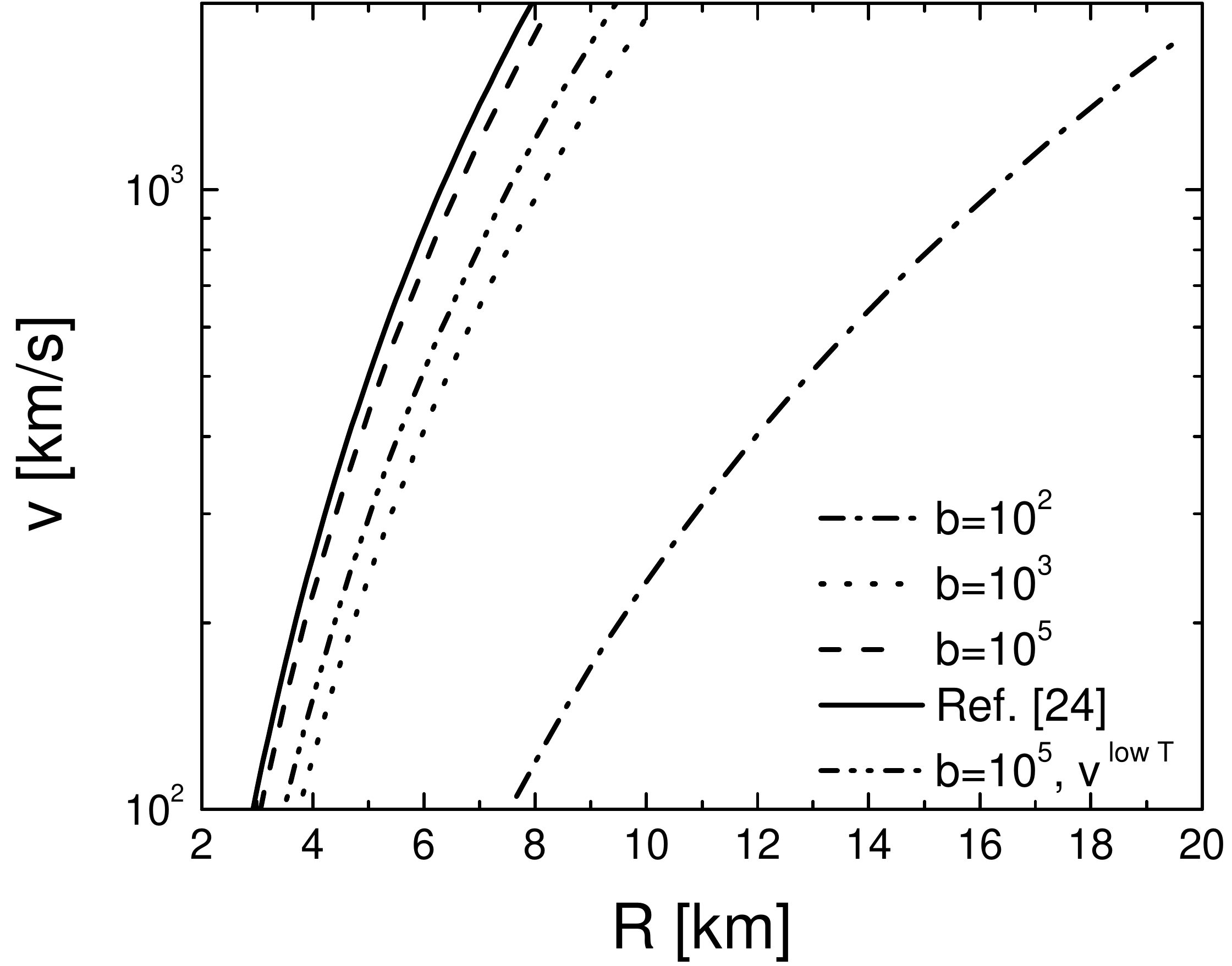}
    \includegraphics[width=0.45\linewidth]{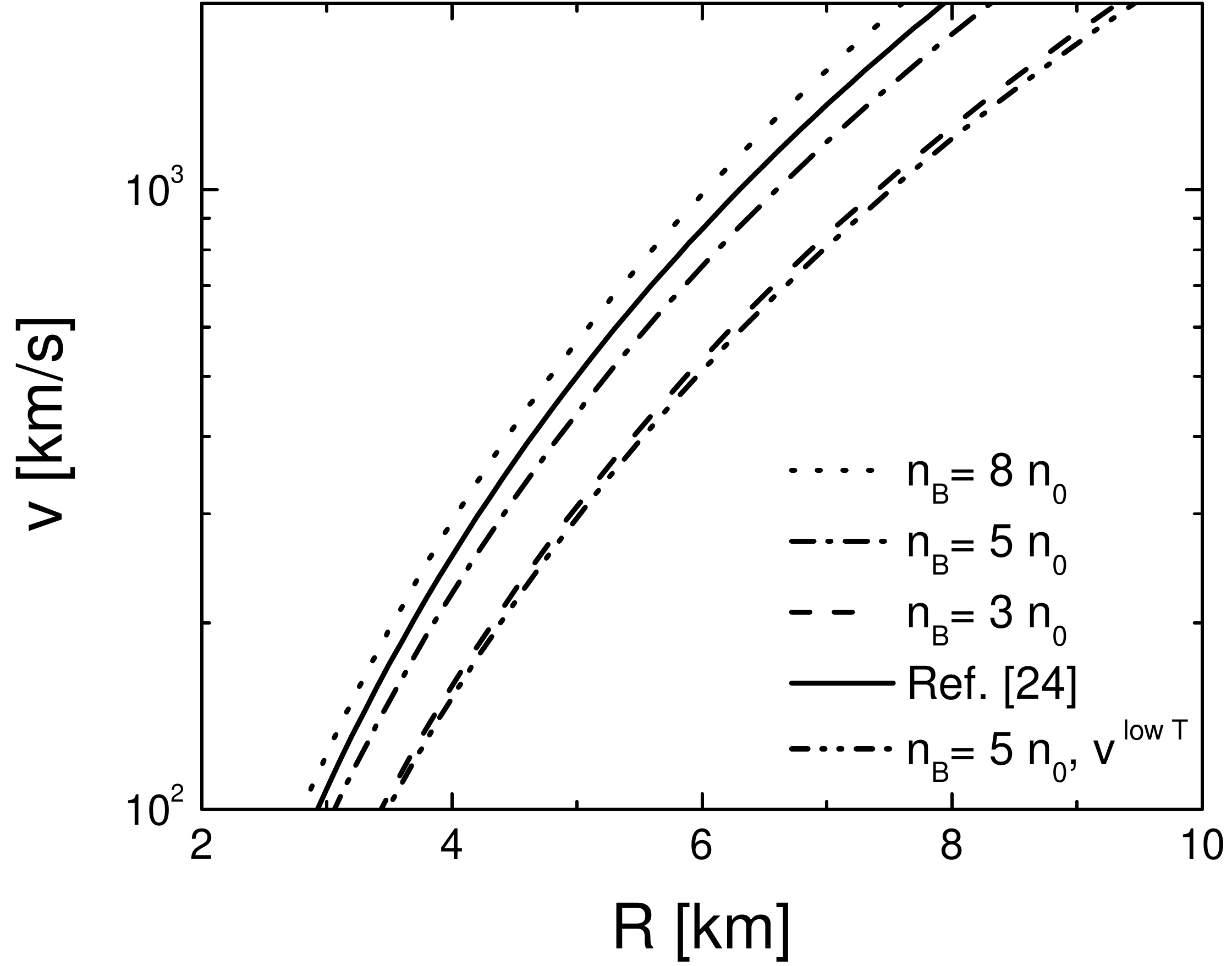}
	\caption{kick velocity as a function of the neutron star radius for different values of the magnetic field and a fixed central density of $n_B = 5n_0$ (left panel) and for different values of the central density at a fixed magnetic field of $b = 10^5$   (right panel).
\label{vR1}}
\end{figure}

\section{Discussion and conclusions}\label{sec6}

In this work we have studied the anisotropic neutrino emission from the core of a NS in the presence of a magnetic field as a possible mechanism to account for the NS's kick velocities. We have modeled the NS core as made out of a magnetized gas of strange quark matter. The neutrino emission is considered imposing stellar equilibrium conditions and accounting for the magnetic field dependence in the chemical potential and temperature of all the thermodynamical quantities involved. A main ingredient is the full numerical calculation both of the polarization and of the heat capacity for low temperatures.

For typical values of densities an temperatures, the inclusion of stellar equilibrium conditions  and the dependence of all thermodynamical quantities on the magnetic field  allow us to obtain a more realistic model to describe the  kick velocity mechanism and cover a wider range of magnetic field values. Our result for the kick velocity tends asymptotically  to the one of Ref.~\cite{Sagert:2007as} where the ideal condition of $\chi=1$ was used, implying very high magnetic fields.

The anisotropic neutrino emission as a source of NS's kick velocities faces many challenges. Ref.~\cite{Kusenko:1998yy} argues that in thermal equilibrium no asymmetry can be produced, even in the presence of parity-violating processes, such as the one considered in the present work. Therefore, the process is more significant during stages where the NS's core is out of equilibrium, such as the very early times after the NS's birth~\cite{Yakovlev:2000jp} or during posterior phase transitions of quark matter inside the core. The problem can be translated to the small neutrino mean free path within the thermalized matter~\cite{Sagert:2007as}. It should be noted however that the magnetic field effect on the interaction rate and thus on the mean free path has not yet been calculated and that, when the magnetic field strength is high, a strong modification of such interaction may be expected (see however Ref.~\cite{Erdas:2002wk}). Studies aiming to incorporate non-trivial effects of magnetic fields have been performed for instance in Ref.~\cite{Joshi:2017vpi, Gorbar:2013uga, Gorbar:2010kc}. Other effects such as the magnetic field induced anisotropic pressures should be consistently accounted for the study of the structural and polarization properties of a strongly magnetized stellar object. In particular, the effect of a smaller longitudinal pressure than the transverse one, produces the appearance of a longitudinal instability of the NS's matter when the magnetic field exceeds some critical value~\cite{Isayev:2012sv,Aurora2003EPJC,PhysRevC.77.015807}. The effects of some of these ingredients are being considered and will be reported  in the near future elsewhere.

\section*{Acknowledgements}

A.P.M. thanks the support of Consejo Nacional de Ciencia y Tecnologia for the sabbatical Grant. No. 264150 and the hospitality of ICN-UNAM during the sabbatical stay, where this work was conceived. Support for this work has been received in part by UNAM-DGAPA-PAPIIT grant numbers IN101515, IN117817,  IA107017 and by Consejo Nacional de Ciencia y Tecnolog\'ia grant number 256494.  A.P.M has been supported by the grant CB0407 and acknowledges receipt of the grant from the Abdus Salam International Centre for Theoretical Physics, Trieste, Italy.  D.M.P. has been also supported by a DGAPA-UNAM fellowship.

%\bibliography{betadecay}

\end{document}